\documentclass[aps,floatfix,amsmath,amssymb]{revtex4}
\usepackage{natbib}
\usepackage{dcolumn}
\usepackage{graphicx}
\usepackage{epstopdf}
\usepackage{color}
\usepackage[usenames,dvipsnames]{xcolor}
\usepackage[normalem]{ulem}
\usepackage{multirow}
\usepackage{ulem}
\usepackage{longtable}

\begin{document}
\title{Effect of inhomogeneities on high precision measurements of cosmological distances}
%
\author{Austin Peel$^{1}$\footnote{austin.peel@utdallas.edu}}
\author{M. A. Troxel$^{1,2}$\footnote{michael.troxel@manchester.ac.uk}}
\author{Mustapha Ishak$^{1}$\footnote{mishak@utdallas.edu}}
\address{$^{1}$Department of Physics, The University of Texas at Dallas, Richardson, TX 75080, USA}
\address{$^{2}$Jodrell Bank Centre for Astrophysics, University of Manchester, Manchester M13 9PL, UK}
\date{\today}
\maketitle
We study effects of inhomogeneities on distance measures in an exact relativistic Swiss-cheese model of the universe, focusing on the distance modulus. The model has $\Lambda$CDM background dynamics, and the `holes' are non-symmetric structures described by the Szekeres metric. The Szekeres exact solution of Einstein's equations, which is inhomogeneous and anisotropic, allows us to capture potentially relevant effects on light propagation due to nontrivial evolution of structures in an exact framework. Light beams traversing a single Szekeres structure in different ways can experience either magnification or demagnification, depending on the particular path. Consistent with expectations, we find a shift in the distance modulus $\mu$ to distant sources due to demagnification when the light beam travels primarily through the void regions of our model. Conversely, beams are magnified when they propagate mainly through the overdense regions of the structures, and we explore a small additional effect due time evolution of the structures. We then study the probability distributions of $\Delta\mu=\mu_\mathrm{\Lambda CDM}-\mu_\mathrm{SC}$ for sources at different redshifts in various Swiss-cheese constructions, where the light beams travel through a large number of randomly oriented Szekeres holes with random impact parameters. We find for $\Delta\mu$  the dispersions $0.004\le \sigma_{\Delta\mu} \le 0.008$ mag for sources with redshifts $1.0\le z \le 1.5$, which are smaller than the intrinsic dispersion of, for example, magnitudes of type Ia supernovae. The shapes of the distributions we obtain for our Swiss-cheese constructions are peculiar in the sense that they are not consistently skewed toward the demagnification side, as they are in analyses of lensing in cosmological simulations. Depending on the source redshift, the distributions for our models can be skewed to either the demagnification or the magnification side, reflecting a limitation of these constructions. This could be the result of requiring the continuity of Einstein's equations throughout the overall spacetime patchwork, which imposes the condition that compensating overdense shells must accompany the underdense void regions in the holes. The possibility to explore other uses of these constructions that could circumvent this limitation and lead to different statistics remains open.
\section{Introduction}\label{intro}
Light from distant objects propagates through a cosmic web of structure on its way to us, and the distance we infer to various such sources helps us build a consistent picture of the expansion history of the universe. In particular, type Ia supernovae observations have become a standard probe of cosmological expansion, providing the most compelling evidence supporting a universe presently dominated by some form of dark energy or a cosmological constant. The Lambda cold dark matter ($\Lambda$CDM) concordance model, which a perturbed Friedmann-Lema\^itre-Robertson-Walker (FLRW) metric describes the spacetime, constitutes our current best description of the cosmic dynamics and is so far consistent with the array of observations we have available. However, the universe is lumpy and inhomogeneous at nearly all scales we observe, and so the $\Lambda$CDM model is assumed to represent a smoothing out of the overall dynamics. The question of precisely how to smooth or average out spacetime inhomogeneities is still open (see \cite{vdH,BR} and references therein), and it is not the subject of this paper.

Varying density environments and local expansion along the path of a light beam distort and magnify (or demagnify) its cross section, which we can use to determine the distance to its source. Different approaches have been employed to account for the biases introduced by intervening structure between sources and observers. For example, one method is via weak gravitational lensing in the context of perturbation theory, where the magnification probability distribution of distant sources can be calculated from the probability distribution of the density contrast \cite{Valageas00}. Luminosity distance fluctuations have been examined in linear theory \cite{BDG}, and effects on the distance-redshift relation due to perturbations at second order have been studied by, e.g., \cite{DMNV,CUMD}. Nonlinear relativistic contributions in weak-lensing convergence have also been recently examined in \cite{ACPUU}. Another approach is the Dyer-Roeder (DR) formula \cite{DR}, which takes into account that the average density along a line of sight will be some fraction of the smoothed background value, thereby modifying the distance-redshift relation due to magnification by a beam partially filled with structure.

Swiss-cheese models provide a third way to incorporate inhomogeneities and have the advantage of retaining nonlinear effects of general relativity that might impact on light propagation. In the general Swiss-cheese approach, spherical regions are removed (holes) from an FLRW background (cheese) and replaced by a different solution of general relativity. Einstein and Straus pioneered this idea using Swarzschild black holes surrounded by vacuum regions \cite{EStraus}. By construction, as long as hole-cheese junctions obey an exact matching and no holes overlap, the dynamics of the background model remains unaffected by the presence of the holes. Note that averaged quantities, such as the expansion rate, however, might be different from that of the background \cite{Buchert11,Sussman11,LRS}. The metric of the entire Swiss-cheese spacetime is also still an exact solution of Einstein's equations, and it can be made statistically homogeneous by an appropriate distribution of the holes. The goal of such a model is to provide a more realistic environment for photons as they traverse the inhomogeneous universe, where most or all of the volume of the holes is underdense; indeed, most lines of sight in the universe are underdense \cite{CEFMUU}. Many authors have studied the effects on distance measures in Swiss-cheese models with varying degrees of realism in the distribution and structure of the holes, both with and without a cosmological constant \cite{MKMR,BTT,Szybka,BF,BN,BC,FKW,VFW,CZ,Valkenburg09,FDU1,FDU2}. It has also been shown recently that the distance-redshift relation in a certain kind of Swiss-cheese model is in fact equivalent to the DR approximation \cite{Fleury14}.

In this work, we study the effects on distance measures in a Swiss-cheese model where spherical regions of constant comoving size are removed from a $\Lambda$CDM background and replaced by the exact inhomogeneous and nonsymmetric metric of Szekeres. Spherically symmetric Lema\^itre-Tolman (LT) holes have often been used in similar investigations with the aim of determining whether we might mistake a nonaccelerating inhomogeneous universe for a homogeneous one with accelerated expansion; for example, see \cite{MKMR,Szybka,BN,BTT}. The Szekeres solution allows us to make the holes anisotropic as well as inhomogeneous, and they therefore evolve differently compared to LT structures, which is important for photons traveling through them \cite{TIP,BCMBMB}. A Szekeres Swiss-cheese construction is still an idealized model of the universe, and it is worth clarifying that we are not advocating it as a realistic representation of the true cosmos. However, we do aim with this model to explore the effects of anisotropic and nonlinear matter clustering that could be important for real light beams.

In Ref. \cite{BC}, the authors studied light propagation in an axially symmetric Swiss-cheese universe with axial light rays. They found that structures of order $\sim$500 Mpc are required in this scenario to explain the apparent dimming of the type Ia supernovae when the background model is Einstein-de Sitter (EdS). Using smaller $\sim$50 Mpc structures did not significantly alter the distance-redshift relation from that of EdS, and including a cosmological constant improved their fits to supernova data in both cases. In this investigation, we use a Swiss-cheese model based on the Szekeres metric, which uses holes with density profiles that are neither axial nor spherical and that have sizes of 30--60 Mpc comoving radius. We allow for random orientations of Szekeres holes and beam paths with random impact parameters. We adopt the viewpoint that the universe is indeed well approximated by the standard $\Lambda$CDM framework, but that nonlinear structures can induce biases in the inference of cosmological distances, and therefore in cosmological parameters, as well. In particular, these effects become increasingly important as we devise observation schemes to probe the universe with ever finer precision.

The layout of the paper is as follows. We first introduce the Szekeres metric in Sec. \ref{sec:Szek_metric}, which we use to build the inhomogeneous holes, and describe how we build the Swiss-cheese model. We then present the formalism of light propagation in an inhomogeneous universe and how we implement it in our code in Sec. \ref{sec:Light_propagation}. In Sec. \ref{sec:Results} we quantify the effect on the distance modulus of sources at various redshifts and with varying degrees of inhomogeneity of the holes. We summarize and conclude in Sec. \ref{sec:Conclusion}.

\section{Building Nonsymmetric Large-Scale Structures in an Exact General Relativistic Framework}\label{sec:Szek_metric}
\subsection{The Szekeres metric}
The Szekeres metric is an exact solution of the Einstein field equations for irrotational dust \cite{Szekeres1,Szekeres2} that can be used to model inhomogeneous and anisotropic structures in the universe. Its general formulation has no Killing vector fields \cite{BST} and therefore no symmetries; however, imposing certain conditions on the metric functions can produce symmetries. Numerous authors have studied Szekeres models in a variety of cosmological and astrophysical contexts, and we refer the reader to the partial list \cite{PK,Bolejko07,BKHC,IRGWNS,BS,NIT,MB,IP,PIT,WH,RS,TPI,IPT,MC} (and references therein) for further details and model-building techniques. We provide only an abridged introduction to the solution and its properties that are relevant to this work.

Two families of the solution exist, which are called class I and class II, and they are distinguished by a different dependence of the metric functions on the coordinates. We use the more general class I family in this work, as it provides a more intuitive framework for building inhomogeneous structures. The class I solution allows for three different types of 2-surface geometries to foliate the three-spaces of constant time. In the parameterization introduced by Hellaby \cite{H1}, this is determined by the metric parameter $\epsilon$, which takes values in the set $\{-1,0,+1\}$. The three cases are called quasihyperbolic, quasiplanar, and quasispherical, with 2-metrics of hyperboloids, planes, and spheres, respectively. We specialize to the quasispherical metric where $\epsilon=+1$ so that we can replace spherical holes excised from a background FLRW spacetime.

The line element in comoving and synchronous coordinates for the quasispherical case is
\begin{equation}
  ds^2=-\mathrm{d}t^2+\frac{(\Phi,_r-\Phi E,_r/E)^2}{1-k}\mathrm{d}r^2+\frac{\Phi^2}{E^2}(\mathrm{d}p^2+\mathrm{d}q^2)\label{eq:Szek_metric}
\end{equation}
in units where $c=1$, and where a comma indicates partial derivative. The purely spatial function $E$ can be written as
\begin{equation}
  E(r,p,q)=\frac{S(r)}{2}\left[\left(\frac{p-P(r)}{S(r)}\right)^2+\left(\frac{q-Q(r)}{S(r)}\right)^2+1\right],\label{eq:E_function}
\end{equation}
where $S(r)$, $P(r)$, and $Q(r)$ are arbitrary functions that control the anisotropy of the dust structure. Setting these three functions to constants specializes the metric to the spherically symmetric LT solution. In that case, the generally nonconcentric $(p,q)$ 2-spheres become concentric, thereby allowing $r$ to be properly viewed as a radial coordinate.

The $p$ and $q$ coordinates vary over $(-\infty,\infty)$, and a coordinate transformation reveals that $p$ and $q$ are stereographically projected coordinates from $\theta$ and $\phi$ angular coordinates on the unit sphere:
\begin{equation}
  p=P(r)+S(r)\cot(\theta/2)\cos\phi,\quad \mathrm{and}\quad q=Q(r)+S(r)\cot(\theta/2)\sin\phi.\label{eq:stereo_proj}
\end{equation}
That is, for each value of $r$, there is a different mapping (in general) of the $p,q$ plane to the unit sphere that depends on the choice of $S$, $P$, and $Q$. By fixing $(t,r)$ and using Eqs. (\ref{eq:E_function}) and (\ref{eq:stereo_proj}) in Eq. (\ref{eq:Szek_metric}), one finds indeed that $E^{-2}(\mathrm{d}p^2+\mathrm{d}q^2)$ is the metric of the unit sphere.

The evolution of the quantity $\Phi(t,r)$, which is often called the generalized scale factor or areal radius, obeys
\begin{equation}
  (\Phi,_t)^2=-k(r)+\frac{2M(r)}{\Phi}+\frac{\Lambda}{3}\Phi^2.\label{eq:Phi_evolution}
\end{equation}
For a given $t$, $\Phi$ is the radius of the comoving sphere labeled by $r$, the area of which is $4\pi\Phi^2$. By Eq. (\ref{eq:Phi_evolution}), it is clear then that each $(p,q)$ 2-sphere evolves independently according to its own Friedmann equation. The function $M(r)$ appears as an arbitrary function of integration and is physically interpreted as the total gravitational mass inside a sphere of constant $r$. A reasonable constraint on $M$ is therefore that it should be non-negative. Finally, $\Lambda$ is the cosmological constant.

Solutions for $\Phi$ can be expressed in closed form when $\Lambda=0$ in terms of a parameter. Three cases are possible depending on the sign of $k(r)$, which determines the type of evolution of the 2-sphere at that $r$. As in FLRW universes, $k>0$ is elliptic, $k=0$ is parabolic, and $k<0$ is hyperbolic. When $\Lambda\ne 0$, analytic solutions for $\Phi$ have been found in terms elliptic functions (see, for example, \cite{BSS}), and we have implemented one such solution in our code in terms of the Weierstrass $\wp$ and associated Jacobi $\vartheta$ functions. A final arbitrary function $t_B(r)$ called the bang time appears in all solutions for $\Phi$ in the combination $(t-t_B(r))$. It arises by integration of Eq. (\ref{eq:Phi_evolution}) and gives the local time of the big bang.

The field equations give one other independent relation, which is the expression for the mass-energy density:
\begin{equation}
  \kappa \rho=\frac{2(M,_r-3M E,_r/E)}{\Phi^2(\Phi,_r-\Phi E,_r/E)},\label{eq:density}
\end{equation}
where $\kappa=8\pi$ in units with $c=G=1$. When $E,_r\neq 0$, the density varies as a dipole around each $(p,q)$ 2-sphere. The density is maximum at a single point $(p_0, q_0)$ for a given $r$ and minimum at the antipodal point, while varying monotonically between.

The metric contains FLRW solutions as special cases, which arise by imposing two conditions on its functions. In this parameterization, we can take the first to be $\Phi(t,r)=a(t)r$, where $a(t)$ is the FLRW scale factor, and the second as $k(r)=k_0 r^2$, where $k_0$ is the curvature index of the FLRW spacetime. It follows also in this limit that $M(r)$ becomes proportional to $r^3$. The functions $S$, $P$, and $Q$ need not assume any particular form, but different choices amount to different coordinate systems on the resulting FLRW spacetime.

\subsection{Description of the Swiss-cheese model}
We construct a model of the universe where spherical regions (holes) are removed from a homogeneous background FLRW metric (cheese) and replaced by inhomogeneous Szekeres structures. We choose the background metric to have $\Lambda$CDM parameters of $(\Omega_\mathrm{m},\Omega_\Lambda)=(0.3,0.7)$ and $H_0=70$ km s${}^{-1}$ Mpc${}^{-1}$. The matching is exact across the spherical junction surfaces, which are comoving hypersurfaces of constant $r=r_\mathrm{match}$, so that the resulting Swiss-cheese spacetime is still an exact solution of Einstein's equations. To achieve the matching, we satisfy the Darmois conditions \cite{BV}, which require that on the hypersurface, the first and second fundamental forms are (independently) equal as evaluated by the metrics on either side. In practice, these conditions imply the same constraints on the metric functions as in the spherically symmetric LT case. They are that (i) the integrated mass of the hole is the same as FLRW region that it replaces; (ii) at the junction, the Szekeres curvature function $k(r)$ becomes $k_0 r^2$, where $k_0$ is the curvature index of the FLRW exterior; (iii) the bang time function $t_B(r)$ becomes constant and equal to the FLRW value (usually 0) at the junction; and (iv) $\Lambda$ has the same value within the holes as in the FLRW background.
\begin{figure}
\centering
  \includegraphics[scale=0.5]{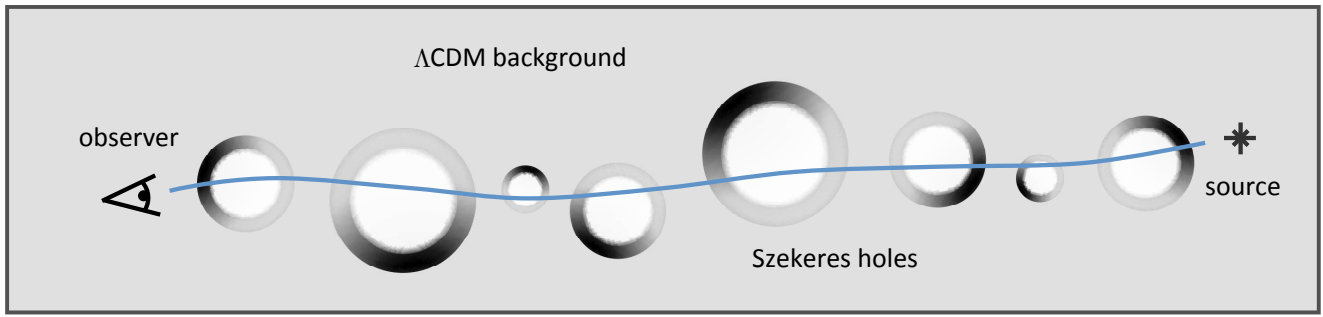}\\
\caption{Schematic of a Szekeres Swiss-cheese model with a light beam propagating to an observer after traversing many inhomogeneous structures. The background is a homogeneous $\Lambda$CDM spacetime, while the holes have an anisotropic density distribution and are built using the Szekeres metric. The central underdense regions of the inhomogeneities have fractional underdensities of $\delta=-0.8$ to mimic real voids in the universe, while the maximum density attained in the overdense compensating shells ranges from about 12 to 24 times the homogeneous background value, as measured today. The holes are to be thought of as distributed throughought the background spacetime, but only those intersected by the beam have been depicted here.}\label{fig:many_holes_diagram}
\end{figure}

In our model, we impose also that the $S$, $P$, and $Q$ Szekeres functions become constant at the junctions as well as for $r>r_\mathrm{match}$, which is not necessary for proper matching, but a convenience for the coordinates to be naturally extended \textit{through} the junction. As mentioned previously, by construction, the dynamics of the background spacetime is not affected by presence of the holes, since the average mass-energy density of the entire spacetime remains unchanged. We therefore do not address the issue of backreaction (see, e.g., \cite{ES,BKS,Buchert00,Buchert01,Rasanen03,LS,Paranjape08,Sussman11,Rasanen11,GW,BR}), which considers effects on cosmic dynamics and parameter determination due to averaging an inhomogeneous spacetime. 

We take as the FLRW background, instead of its more usual metric form, the Szekeres solution specialized to its FLRW limit. This is more convenient for our code, since then we can work consistently within the Szekeres framework and do not have to change between metrics. The above matching conditions then imply that $k(r)$ and $t_B(r)$ take on the appropriate forms not only at the junction surface, but also beyond for $r>r_\mathrm{match}$. $M(r)$ becomes proportional to $r^3$ in the FLRW limit, as noted above, and our parameterization of $M$ makes it straightforward to achieve this (see Eq. (\ref{eq:M}) below).

We are interested ultimately in light propagation through many randomly placed holes, each with a random impact parameter, but we do not construct the entire spacetime from the outset. We achieve the effect of an observer receiving a bundle of light rays that has propagated through many intervening holes by patching together alternating Szekeres and FLRW regions as we integrate along the path of the light. This is sufficient for analyzing the effects on redshift and distance for a single source, and by repeating the process for different realizations containing different distributions of holes, we simulate the effect of having many different lines of sight in a $\Lambda$CDM universe filled with Szekeres holes. An important implicit assumption here is that the different lines of sight are uncorrelated.

Figure \ref{fig:many_holes_diagram} gives a schematic illustration of our Swiss-cheese model setup. For simplicity, we take the holes at first to all have the same size (approximately 30 Mpc radius) and density distribution but different orientations and impact parameters along a line of sight. Adjacent Szekeres regions also do not touch, instead having $\Lambda$CDM spaces of approximately $5$ Mpc between them, although this is not a requirement of the framework. The holes are therefore rather tightly packed, and the close overdense shells of neighboring holes are meant to mimic the filamentary structure of the cosmic web. We later test the effect of multiple hole sizes and strengths of the anisotropy (see the end of Sec. \ref{subsec:multiple_Szek}).

\subsection{Building the Szekeres holes}\label{sec:model_building}
To build the Szekeres structures that serve as the holes in the $\Lambda$CDM spacetime, we start by designing the desired density profile at the present time $t_0$, which in our units is $t_0=4.13$ Gpc (equivalently, $\sim$13.5 Gyr) in the background. The initial structures have radii of about $30$ Mpc, and the interior void regions have fractional underdensities of $\delta=-0.8$. We have chosen this size for the structures, as well as for their interiors not to be completely empty ($\delta=-1$), in order to represent typical voids in the real universe (see, e.g., \cite{RQP,CJ,SLHWWW}). Throughout the model we set the bang time function $t_B=0$, and the fact that $t_B,_r=0$ is equivalent to only considering growing modes in these models. It is conceptually simpler to consider first the spherically symmetric (LT) density and then introduce anisotropy via the $S$, $P$, and $Q$ functions---or, more importantly, their derivatives---which have the effect of redistributing the dust within shells. In particular, the structures we end up with are underdense and homogeneous in their interiors and have compensating mass shells of thickness about $8$ Mpc. The matter in the outer shells is distributed anisotropically, which we achieve with $Q,_r\neq 0$ within this range of $r$; see \cite{TPI} for another example of this approach.

The spherically symmetric density is given by
\begin{equation}
  \rho_{LT}(t_0,r)=\rho_{\mathrm{bg},0}\left(1-A\,e^{-b\,(r/\sigma)^n}\left[1-\frac{b\,n}{3} \left(\frac{r}{\sigma}\right)^n\right]\right),\label{eq:rho_LT}
\end{equation}
where we use the parameter set $(n,b,\sigma,A)=(10,0.8,0.025,0.8)$, and $\rho_{\mathrm{bg},0}=(3/8\pi)\Omega_\mathrm{m} H_0^2$ indicates the present background density. This parameterization is a modified version of one we and other authors have used previously \cite{PIT,Bolejko07}. The additional parameter $A$ allows the structure to have $\delta=-0.8$ in the underdense central region while still satisfying the matching conditions. The term in square brackets of Eq. (\ref{eq:rho_LT}) allows for a simple resulting form of the associated $M$ function, which we obtain next. 
\begin{figure}
\centering
\begin{tabular}{c c}
  \includegraphics[scale=0.68]{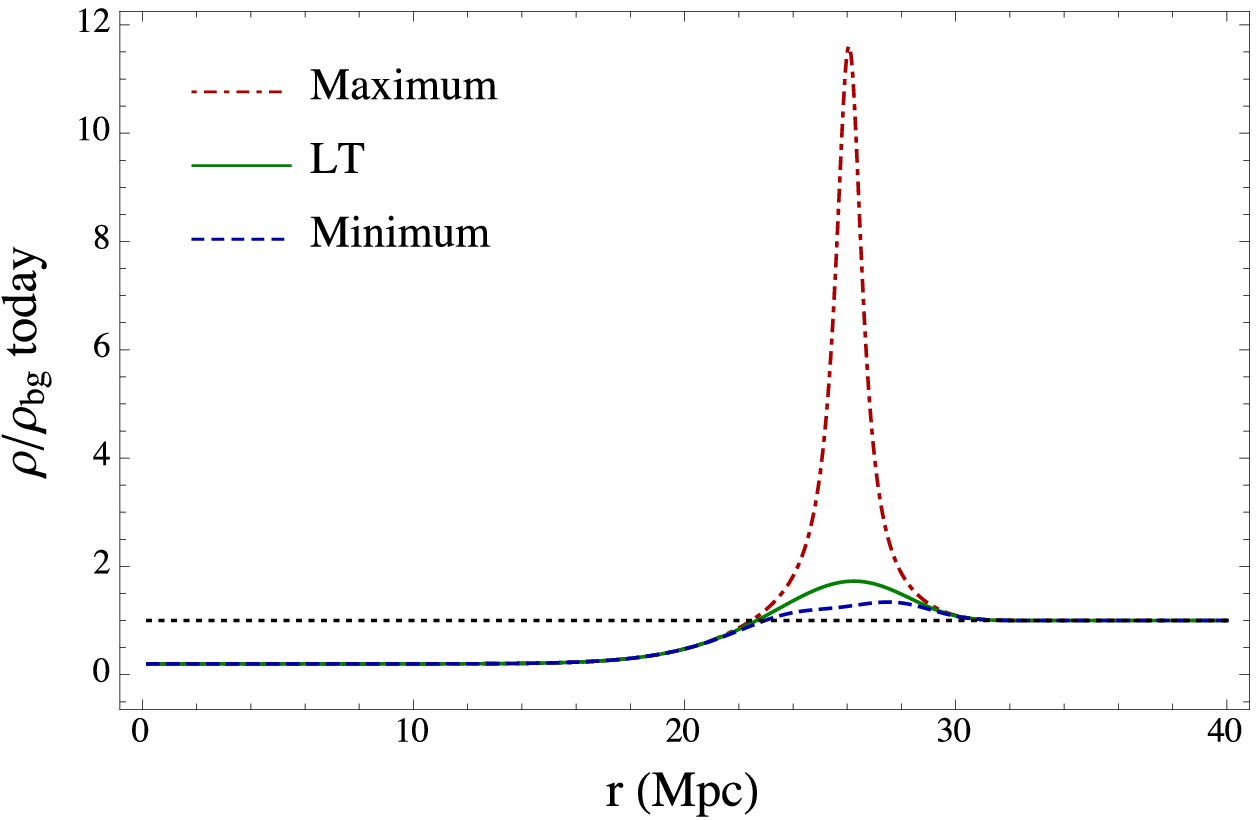}\label{fig:density_today}&
  \includegraphics[scale=0.48]{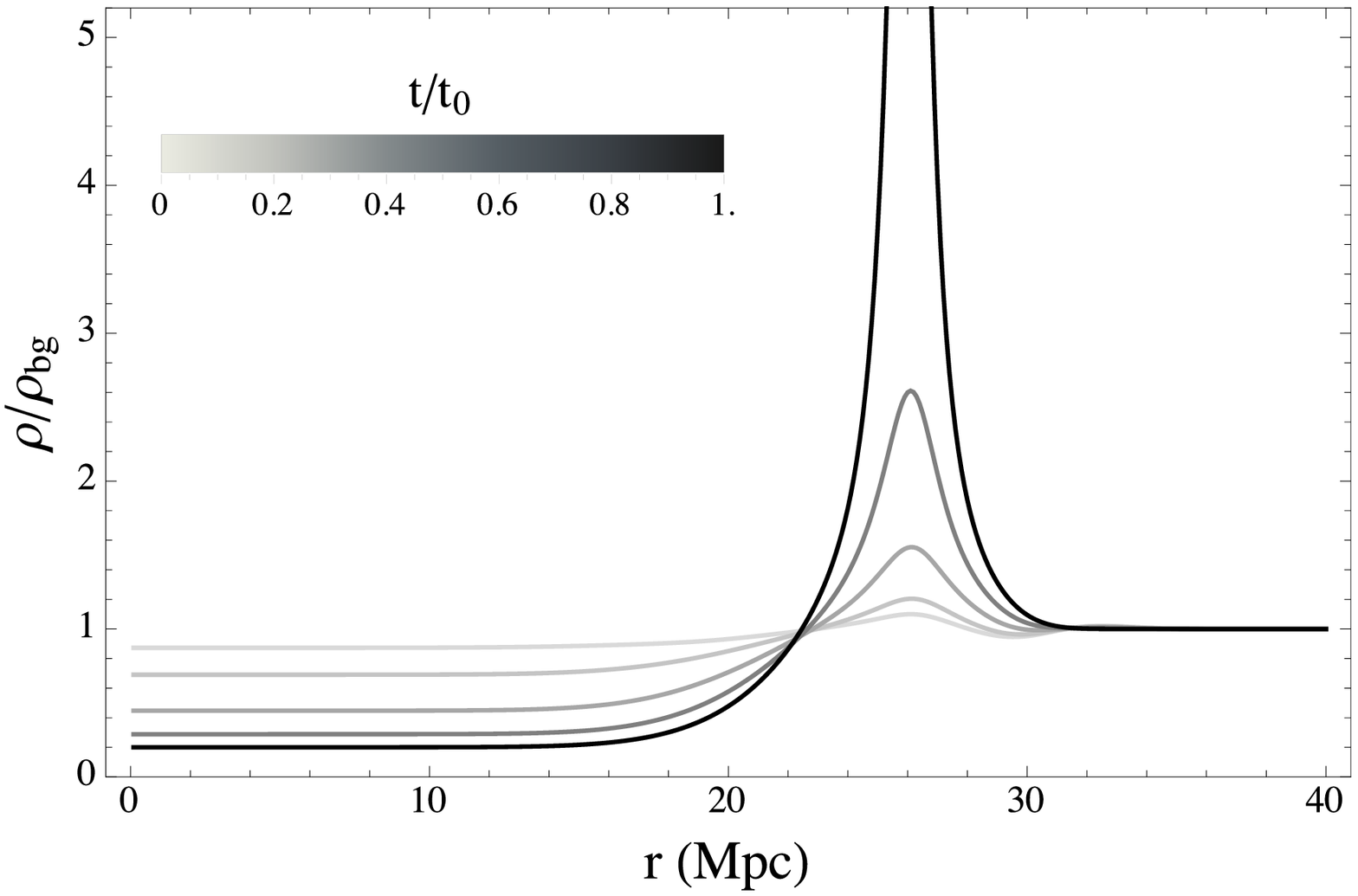}\label{fig:density_evolution}\\
\end{tabular}
\caption{LEFT: Density profile of the 30 Mpc compensated void structures today in units of the $\Lambda$CDM background density. The top (dash-dotted red) curve shows the maximum density attained at each value of $r$, while the bottom (dashed blue) shows the minimum. The middle (solid green) curve is the density profile of the corresponding spherically symmetric LT model, and a dotted line at $\rho/\rho_{\mathrm{bg}}=1$ is shown for reference. The peak of the maximum density is $\rho/\rho_{\mathrm{bg},0}\approx 12$ and occurs at around 26 Mpc. RIGHT: Evolution of the density profile along the direction containing the maximum density peak. The darkest curve is at $t_0$, and the progression toward lighter curves is the same profile at successively earlier times, the lightest being $t/t_0=0.01$.}\label{fig:density}
\end{figure}

The resulting mass function $M(r)$ is obtained by integrating Eq. (\ref{eq:density}) with $E,_r$ set to zero and applying the gauge freedom in the choice of $r$ coordinate: $\Phi(t_0,r)=r$. The expression becomes
\begin{align}
  M(r)&=4\pi\int^r_0 \rho_{LT}(t_0,x)\,x^2~\mathrm{d}x\nonumber\\
  &=\frac{4\pi}{3}\rho_{\mathrm{bg},0}\,r^3\left(1-A e^{-b(r/\sigma)^n}\right).\label{eq:M}
\end{align}
In the LT metric, these same profiles would describe a spherically symmetric structure with a maximum overdensity in the compensating shell of $\rho/\rho_{\mathrm{bg},0}\approx1.7$. To introduce anisotropy, we choose the additional Szekeres $S$, $P$, and $Q$, functions to be
\begin{equation}
  S(r)=0.4,\qquad P(r)=0,\quad\mathrm{and}\qquad Q(r)=\frac{A_Q}{2}\left[1-\tanh\left(\frac{r-b_Q}{\sigma_Q}\right)\right],\label{eq:SPQ}
\end{equation}
where $(A_Q, b_Q, \sigma_Q)=(70, 26, 2.5)$ Mpc. Physically, these functions produce an overall density dipole-like structure, where mass from one side of the overdense shell has been transplanted to the opposite side. The maximum value of the overdensity is now $\rho/\rho_{\mathrm{bg},0}\approx 12$, while the opposite side has been reduced to $\rho/\rho_{\mathrm{bg},0}\approx 1.3$. Due to constraints of the matching and the nature of the metric functions, the peak in the overdensity cannot be made arbitrarily large without inducing shell crossings. We have therefore chosen a moderate value to work with at first and later test the impact of augmenting it (see Sec. \ref{subsec:multiple_Szek}). Such values are also not unrealistic, as simulations of structure formation in a $\Lambda$CDM universe have shown that overdensities within filaments can range from between a few times to up to 25 times the mean density \cite{CKC}; see also Fig. 3 of \cite{BF}, which plots the density contrast along a random line of sight through the Millennium simulation \cite{Springel05,Boylan09}. Within the overdense shell, there is a monotonic and smooth decrease in the density from the maximum point to the minimum point. The left panel of Fig. \ref{fig:density} plots the present density profile of the holes along lines encountering the extreme values of $\rho$ (in units of the background) at each $r$. The maximum is dash-dotted (red), the minimum is dashed (blue), and the corresponding spherically symmetric LT curve is solid (green).

An exact matching across the comoving $r_\mathrm{match}=35$ Mpc surface would generally involve two different metrics, which in our case is Szekeres within the holes and FLRW in the background. However, since the Szekeres metric reduces to FLRW in limiting cases of the metric functions, we simply allow the functions to take on their FLRW values beyond $r_\mathrm{match}$, and therefore the entire spacetime is in fact Szekeres. In practice, our functions  have an abrupt but smooth transition between the holes and the background, instead of using piecewise functions patched together at exactly $r=r_\mathrm{match}$. The density, for example, is different from the would-be FLRW background density by only $\sim$10${}^{-7}$ percent at $r=35$ Mpc, and decreases rapidly with increasing $r$. We have verified that using a smooth transition instead of a sharp piecewise one does not affect our results.

The right panel of Fig. \ref{fig:density} shows the evolution of the density profile along a curve through the maximum density at each value of $r$. Starting from $t_0$ (darkest curve) and looking at progressively earlier times, the density peak at $r\approx 26$ Mpc decreases toward the background value $\rho_{\mathrm{bg}}$, while the underdense central region increases toward it. By $t/t_0=0.01$ (lightest curve), the structure has been reduced to only a slight deviation from the homogeneous background. In the future, the model encounters a shell crossing singularity by $t/t_0\approx 1.2$, since there is no pressure in these models to avoid such a nonphysical divergence in the density. Despite some caveats, like the absence of pressure and rotation, we expect the models to capture the density environments encountered by typical light beams that pass through mostly underdense, and occasionally overdense, regions of the universe before reaching us.

Szekeres models in general have 6 functional degrees of freedom in the choice of $\{M(r),k(r),t_B(r),S(r),P(r),Q(r)\}$, plus a gauge freedom to choose a new $r$ coordinate. By redefining $r$ in terms of one of these functions (e.g., $r'=M(r)$), the Szekeres model is completely determined, and its evolution can be solved. In this work, we have chosen explicit forms for $\{M(r),t_B(r),S(r),P(r),Q(r)\}$ along with $r\rightarrow r'=\Phi(t_0,r)$ (dropping primes for convenience) and therefore solve for $k(r)$ by the following integral,
\begin{equation}
  t-t_B(r)=\int^\Phi_0 \frac{\mathrm{d}x}{\sqrt{-k(r)+\frac{2M(r)}{x}+\frac{\Lambda}{3}x^2}}.\label{eq:k_integral}
\end{equation}
Equation (\ref{eq:k_integral}) arises from integrating Eq. (\ref{eq:Phi_evolution}) with respect to $t$, which has become a constraint equation on $k$ by our choice of how to use the degrees of freedom. Solving for $k$ is carried out in our code semi-analytically using the Carlson symmetric forms of elliptic integrals \cite{Carlson88,Carlson95}.

\section{Light Propagation in an Inhomogeneous Universe}\label{sec:Light_propagation}
Determining the path of a light ray through a general spacetime requires solving the null geodesic equations for the metric. Notions of distance typically used in cosmology, including the angular diameter distance $d_A$ and luminosity distance $d_L$, further require knowledge of how neighboring geodesics spread or converge relative to each other along their parameterized paths. In other words, we must study light beams, or bundles of light rays, in addition to single light rays in order to connect geodesics to physically meaningful measures of distance. The Sachs formalism provides a framework for this, where the evolution of the beam cross section, which gives the distance, is related to Ricci and Weyl focusing terms. For more details of these results and applications or extensions of the formalism, see \cite{Perlick04,SEF,FDU1,TIP,GMNV,FGMV}.

\subsection{Sachs formalism}\label{sec:Sachs_formalism}
To develop the formalism, we first consider a fiducial null geodesic $\gamma$, affinely parameterized by $\lambda$, and with tangent vector components $k^\mu=\mathrm{d}x^\mu/\mathrm{d}\lambda$. We imagine $\gamma$ to belong to an infinitesimal bundle of geodesics that are all parameterized by $\lambda$. Neighboring geodesics to $\gamma$ (an infinitesimal distance away) can be described by a separation vector whose components are $\xi^\mu=\mathrm{d}x^\mu/\mathrm{d}\sigma$, where $\sigma$ is a parameter that labels the family of neighboring geodesics. Distance measures involve knowing how the shape or cross-sectional area of light beams evolve along the geodesic, so we expect that knowing how $\xi^\mu$ changes with $\lambda$ will provide this information. It can be shown that $\xi^\mu$ obeys
\begin{equation}
  \frac{D^2\xi^\mu}{d\lambda^2}=R^\mu{}_{\nu\alpha\beta}\,k^\nu\,k^\alpha\,\xi^\beta,\label{eq:geo_dev}
\end{equation}
where $D/d\lambda=k^\mu\nabla_\mu$, or the covariant derivative along $k$, and $R^\mu{}_{\nu\alpha\beta}$ is the Riemann curvature tensor. Equation (\ref{eq:geo_dev}) is the well-known geodesic deviation equation. 

For an observer with 4-velocity components $u^\mu$ observing the light ray, we can set up a tetrad of vectors called the Sachs basis consisting of $u^\mu$, $k^\mu$, $(E_1){}^\mu$, and $(E_2){}^\mu$. Vectors $(E_1){}^\mu$ and $(E_2){}^\mu$ are space-like and span the 2-space screen orthogonal to both $u^\mu$ and $k^\mu$; they therefore comprise a 2D basis on which to project $\xi^\mu$. It is only this projection that is relevant to the observer, since his line of sight is directed along $k^\mu$. The screen basis vectors are orthonormal as well as parallely propagated along the geodesic; that is,
\begin{equation}
  (E_a){}^\mu (E_b){}_\mu=\delta_{ab},\qquad \mathrm{and}\qquad k^\mu\nabla_\mu(E_a){}^\nu=0,
\end{equation}
where $a,b\in \{1,2\}$, and $\delta_{ab}$ is the Kronecker delta.

We can write the decomposition of $\xi$ relative to the Sachs basis as
\begin{equation}
  \xi=\xi^1 E_1+\xi^2 E_2 +\xi^k k,\label{eq:xi_decomp}
\end{equation}
where $\xi^1=\xi^\mu (E_1)_\mu$ is the component of $\xi$ along $E_1$, and so on. There is no component along the observer's $u$ vector due to the orthogonality relations of the basis set and the fact that $\xi^\mu k_\mu=0$. This can be seen by taking the inner product of $\xi$ with $k$ in Eq. (\ref{eq:xi_decomp}).

We restrict our attention then to the components $\xi^a$ that lie within the screen, since they characterize the cross section of the beam viewed by the observer. Projecting Eq. (\ref{eq:geo_dev}) onto the screen basis gives the evolution of $\xi^a$ along the path of the beam:
\begin{equation}
  \frac{d^2\xi^a}{d\lambda^2}=\mathcal{T}^a{}_b\,\xi^b,\label{eq:Jacobi}
\end{equation}
where $\mathcal{T}{}_{ab}=R_{\mu\nu\alpha\beta}\,(E_a)^\mu\,k^\nu\,k^\alpha\,(E_b)^\beta$ is known as the optical tidal matrix. We note that index placement on $\mathcal{T}$ and similar quantities is not important, except to keep with standard summation notation, since $a$ and $b$ are raised and lowered by $\delta_{ab}$. The four components that comprise $\mathcal{T}{}_{ab}$ can be written in terms of the Ricci and Weyl curvatures of the spacetime as follows.
\begin{equation}
  \mathcal{T}=\begin{pmatrix} \,\mathcal{R}-\mathrm{Re}\,\mathcal{F} & \mathrm{Im}\,\mathcal{F} \\[0.3em] \mathrm{Im}\,\mathcal{F} & \mathcal{R}+\mathrm{Re}\,\mathcal{F}\, \end{pmatrix},\label{eq:opt_tid_matrix}
\end{equation}
where
\begin{equation}
  \mathcal{R}=-\frac{1}{2}R_{\mu\nu}\,k^\mu\,k^\nu
\end{equation}
is the Ricci focusing term, and
\begin{equation}
  \mathcal{F}=-\frac{1}{2}C_{\alpha\beta\mu\nu}\,\epsilon^\alpha\,k^\beta\,\epsilon^\mu\,k^\nu
\end{equation}
is the Weyl focusing term. We have introduced a complex form of the screen basis with $\epsilon=E_1-iE_2$, and $R_{\mu\nu}$ is the Ricci tensor, while $C_{\alpha\beta\mu\nu}$ is the Weyl tensor.

As Eq. (\ref{eq:Jacobi}) is linear, there exist 2-matrices that take the initial conditions (i.e., $(\xi^a)_0=\xi^a{}|_{\lambda=0}$ and $(d\,\xi^a/d\lambda)_0=d\,\xi^a/d\lambda{}|_{\lambda=0}$ at the observation event) to the solution $\xi^a(\lambda)$ at some event farther along the geodesic in a linear way. Noting that $(\xi^a)_0=0$ at the observer, since the beam converges there, we can then write
\begin{equation}
  \xi^a(\lambda)=\mathcal{D}^a{}_b\,\left(\frac{d\,\xi^b}{d\lambda}\right)_0,
\end{equation}
where $\mathcal{D}$ is known as the Jacobi matrix. By Eq. (\ref{eq:Jacobi}), this implies
\begin{equation}
  \frac{d^2\,\mathcal{D}^a{}_b}{d\lambda^2}=\mathcal{T}^a{}_c\,\mathcal{D}^c{}_b,
\end{equation}
with the initial conditions
\begin{equation}
  \mathcal{D}_0=\begin{pmatrix} \,0 & 0\,\, \\[0.3em] \,0 & 0\,\,\end{pmatrix}\,\qquad\mathrm{and}\qquad\left(\frac{d\,\mathcal{D}}{d\lambda}\right)_0=\begin{pmatrix} \,1 & 0\,\, \\[0.3em] \,0 & 1\,\,\end{pmatrix}
\end{equation}
at $\lambda=0$. It is the elements of the $\mathcal{D}$ matrix, solved at some value of the parameter $\lambda$, that determine the angular diameter distance to that event along the null geodesic. It turns out that the relation to $d_A$ is simply
\begin{equation}
  d_A=\sqrt{|\det{\mathcal{D}}|}.\label{eq:and_diam_dist}
\end{equation}

For certain cosmological observations like supernovae, the luminosity distance $d_L$ and associated distance modulus $\mu$ are more practical. The connection between $d_A$ and $d_L$ is provided by the distance duality relation
\begin{equation}
  d_L=(1+z)^2 d_A,
\end{equation}
which holds in any spacetime where photon number is conserved and is based on the Etherington reciprocity relation \cite{Etherington}. The distance modulus is defined as
\begin{equation}
  \mu=5\log_{10}\left(\frac{d_L}{\mathrm{Mpc}}\right)+25,\label{eq:mu}
\end{equation}
and the redshift in the geometric optic approximation is determined by the ratio of the light's wavelength at observation to its wavelength at emission; that is, 
\begin{equation}
  1+z=\frac{(k^\mu u_\mu)_\mathrm{e}}{(k^\mu u_\mu)_\mathrm{o}},
\end{equation}
where the subscripts o and e refer to observation and emission, respectively.

The lensing properties of beams in a Swiss-cheese model can be compared to those of the smooth background model by defining the magnification matrix $\mathcal{A}$ as the Jacobi matrix scaled by $D_A$, the angular diameter distance to a source if the light had propagated solely in the background spacetime: $\mathcal{A}=\mathcal{D}/D_A$. Then $\mathcal{A}$ can be decomposed into a rotation matrix and a distortion matrix, giving
\begin{equation}
  \mathcal{A}=\left(\,\begin{matrix}\cos\omega & \sin\omega \\ -\sin\omega & \cos\omega\end{matrix}\,\right)
  \left(\,\begin{matrix}1-\kappa-\gamma_1 & -\gamma_2 \\ -\gamma_2 & 1-\kappa+\gamma_1\end{matrix}\,\right),
\end{equation}
where then the convergence $\kappa$ and shear $\gamma=\gamma_1+\mathrm{i}\gamma_2$ can be expressed in terms of matrix elements as
\begin{align}
  \kappa&=1-\frac{A_{11}+A_{22}}{2\cos\omega},\label{eq:lensing_kappa}\\
  \gamma_1&=\frac{1}{2}\left[(A_{22}-A_{11})\cos\omega+(A_{12}+A_{21})\sin\omega\right],\label{eq:lensing_gamma}\\
  \gamma_2&=\frac{1}{2}\left[(A_{22}-A_{11})\sin\omega-(A_{12}+A_{21})\cos\omega\right],
\end{align}
and where the rotation angle is
\begin{equation}
  \omega=\arctan\left(\frac{A_{12}-A_{21}}{A_{11}+A_{22}}\right).
\end{equation}
The rotation term arises physically due to multiple shearings of the beam by successive lenses \footnote{We thank P. Fleury for bringing this point to our attention.}, but for the models considered in this work, we find that $\omega$ is negligible. Defining further the magnification $m=(\mathrm{det}\mathcal{A})^{-1}$ in the usual way---denoted here as $m$ to avoid confusion with the distance modulus---it is straightforward to show that
\begin{equation}
  m=\frac{1}{(1-\kappa)^2-|\gamma|^2+\omega^2}=\left(\frac{D_A}{d_A}\right)^2.
\end{equation}
In other words, magnification here refers to the observed size of a source in our Swiss-cheese model compared to that of a pure FLRW model.

\subsection{Code implementation}
To solve the evolution and light propagation equations in Szekeres spacetimes, we have implemented a specific semi-analytic calculation using a combination Fortran and Mathematica code package developed by the present authors and others. The initial work in solving the geodesic equations and luminosity distance for observers at the origin (in terms of partial derivatives of the null vector components) for general Szekeres models was carried out in \cite{NIT}. Subsequent works \cite{TIP,TPI,PIT} used code developed to extended the initial approach and accommodate observers at arbitrary positions in and around a Szekeres structure. The code package solves the model evolution via elliptic functions and light propagation by the Sachs formalism as described in Sec. \ref{sec:Sachs_formalism}. For calculations in the present work, the latter code has been modified to meet the needs of a Swiss-cheese model, which includes allowing for multiple randomly oriented structures along a line of sight.

The strategy for obtaining $d_A$ in our Swiss-cheese model is to propagate a light beam from observer to source by solving the geodesic and Jacobi evolution equations simultaneously. In reality, the beam travels from source to observer, but our calculations proceed the other direction, since we want to take the observation event, where the beam is converged to a vertex, as boundary condition. We start with a single coordinate patch that contains a Szekeres inhomogeneous structure within $0\leq r\leq r_\mathrm{match}$. Beyond $r_\mathrm{match}$, the spacetime is $\Lambda$CDM, and we place the observer  in this outer region. The beam is propagated across the structure, where it encounters first the homogeneous background, then inhomogeneous structure, and then it returns to the homogeneous space. Once back in the $\Lambda$CDM background, the beam transitions to a new coordinate patch of the same type, but with a different orientation of the structure and a new impact parameter. The process is repeated until the desired redshift is reached, whereby $d_A$ is determined using Eq. (\ref{eq:and_diam_dist}) with the $\mathcal{D}$ matrix elements at the source.

We first choose the coordinates of the observation event, which are $t=t_0$, $r=r_0>r_\mathrm{match}$, and $p_0$ and $q_0$ are given random values. In practice, we randomize the angular coordinates $\theta\in(0,\pi\rbrack$ and $\phi\in\lbrack0,2\pi)$ and convert them to $(p,q)$ using Eq. (\ref{eq:stereo_proj}) with $r=r_0$. This is more convenient, as the angular coordinates have finite ranges, and their location relative to the structure is more intuitive. We also note that 0 is excluded from the range of $\theta$ due to the singularity of the coordinate transformation there. As the geodesics are past directed, the parameter decreases toward zero as $t\rightarrow t_0$. The choice of $r_0$ therefore influences at what redshift the beam encounters the final structure along the line of sight before reaching the observer. Larger $r_0$ corresponds to greater distance between the observer and the inhomogeneity surrounding the spatial coordinate origin $r=0$, and therefore translates to a higher redshift of the final hole.

The geodesic is aimed toward the structure by a choice of $(k^\mu)_0=k^\mu|_{\lambda=0}$, for which we are free to take $(k^t){}_0=-1$ by a suitable choice of parameter and to enforce $z=0$ at the observer. To achieve a randomly chosen impact parameter $b_i$, for which the maximum is taken to be at the edge of the structure, $(k^p){}_0$ and $(k^q){}_0$ are randomized within bounds set by the geometry of the problem. Again, in practice we work first with the angular $(k^\theta)_0$ and $(k^\phi)_0$ at the fixed $r_0$ and then convert to $(k^p){}_0$ and $(k^q){}_0$ after. Once $b_i$ is chosen, the maximum range of $(k^\theta)_0$ is determined by assuming temporarily $(k^\phi)_0=0$. $(k^\theta)_0$ is then given a random value within this range, and then we solve for $(k^\phi)_0$ by the constraint imposed by $b_i$. Finally, $(k^r){}_0$ is obtained by solving $k^\mu\,k_\mu=0$, which is the condition that the geodesic be null.

We conclude this section by pointing out another useful way to view the process of light propagating through our Swiss-cheese model that only involves a single coordinate patch. Consider the observer at $(x^\mu)_0$ in the $\Lambda$CDM region, so $r_0>r_\mathrm{match}$. Solve the geodesic and Jacobi evolution equations backward in time across the hole until the beam reaches the $r_0$ sphere on the opposite side of the structure. Rotate to a new $(\theta,\phi)$ position on the $r=r_0$ sphere, aim the light beam back toward the structure, and integrate across it again. The beam crosses the structure now along a different path from previously and with a different impact parameter, where $t$ decreases and $\lambda$ increases continually. Iterate this process, letting final values from one hole serve as inputs for the next, until the desired redshift of the source is reached.

\begin{figure}
\centering
  \includegraphics[scale=0.5]{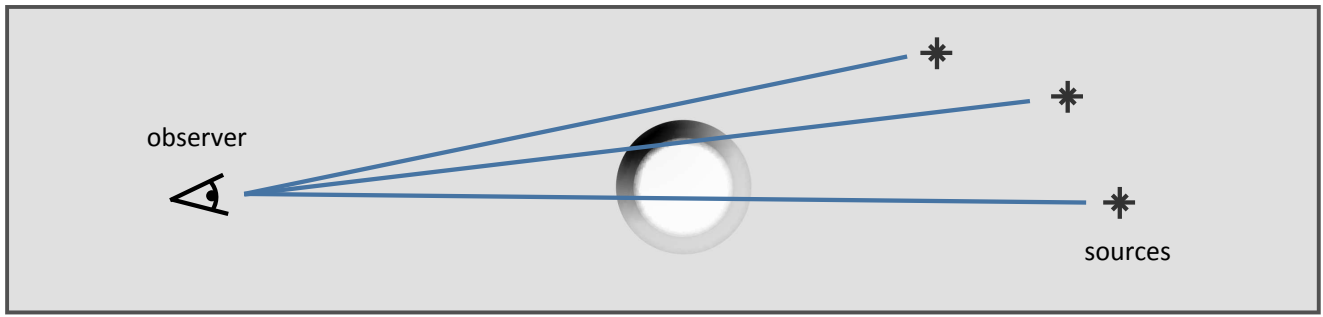}\\
\caption{Illustration of a few possible beam paths through one hole. The beam may be almost `radial', spending most of its time in the underdense central region and passing through the overdense shell in two places. Otherwise, it can pass near the edge through just the overdense region, or even miss the structure entirely, depending on initial conditions. Bending of the beams due to lensing has not been depicted.}\label{fig:one_hole_diagram}
\end{figure}

\section{Results}\label{sec:Results}
\subsection{Single hole: particular paths through a Szekeres structure}
As a first test of our theoretical approach and numerical code, we verify that the results are what we expect for light propagating along particular paths through a single hole. With the observer in the $\Lambda$CDM region, a light beam can travel many paths across a structure placed at some redshift $z>0$, including the extreme cases of radially through the center, or alternatively missing the structure altogether. See Fig. \ref{fig:one_hole_diagram} for an illustration of a few such possible paths. With anisotropic structures like the ones we use, a perfectly radial trajectory is not actually possible, since an initially radial geodesic will not remain so as it traverses the inhomogeneity. (The notion of radial is not even well defined in the general Szekeres metric: $r$ is not truly a radial coordinate, and the coordinate origin need not reside at the geometric center of the ($p,q$) 2-spheres.) However, we do not expect the bending of the geodesics to be significant for our structures, and so by a `radial' path we will mean one that only deviates slightly from a line connecting two antipodal points on the 2-sphere defined by $r=r_\mathrm{match}$. In this case, the photons spend most of their time in the underdense central region of the structure and minimal time in the overdense shell.

For beams that do not cross the structure and instead propagate only in the background spacetime, we find no deviation in $d_A$ from the $\Lambda$CDM result. This is an important confirmation, since we are not simply solving the usual integral for $d_A$ in a $\Lambda$CDM universe, but rather using an entirely Szekeres framework specialized to its FLRW limit. We are solving the full set of geodesic and $\mathcal{D}$ matrix evolution equations, and the result is not sensitive to initial conditions $(x^\mu)_0$ and $(k^\mu)_0$, so long as they are chosen so that the beam avoids the structure.

Beams that cross the structure can experience quite varied density and local expansion environments, depending on the orientation of the hole and the beam's impact parameter. For example, the beam might graze just the overdense shell, or it could travel closer along the axis of the overall dipole, where it would encounter both the maximum and minimum (over-)density values in the mass-compensating shell. As with spherically symmetric LT structures, we do not expect an exact cancellation effect in $d_A$ for a typical beam path, even when the path is `radial'. This is because holes built using either metric evolve during the time it takes the light to cross them, so the photons experience a nonzero integrated density contrast passing from one side to the other. We might expect, however, certain beam paths across Szekeres structures to result in different values of $d_A$ than are possible in their corresponding LT structures. In addition to anisotropic density distributions that have locations of larger maximum $\rho$, Szekeres structures have been shown to have gravitational infall that is faster than that in their spherically symmetric counterparts \cite{IP,PIT,Bolejko07,TPI}, and this could have a noticeable effect on the distance.
\begin{figure}
\centering
\begin{tabular}{c c}
  \includegraphics[scale=0.6]{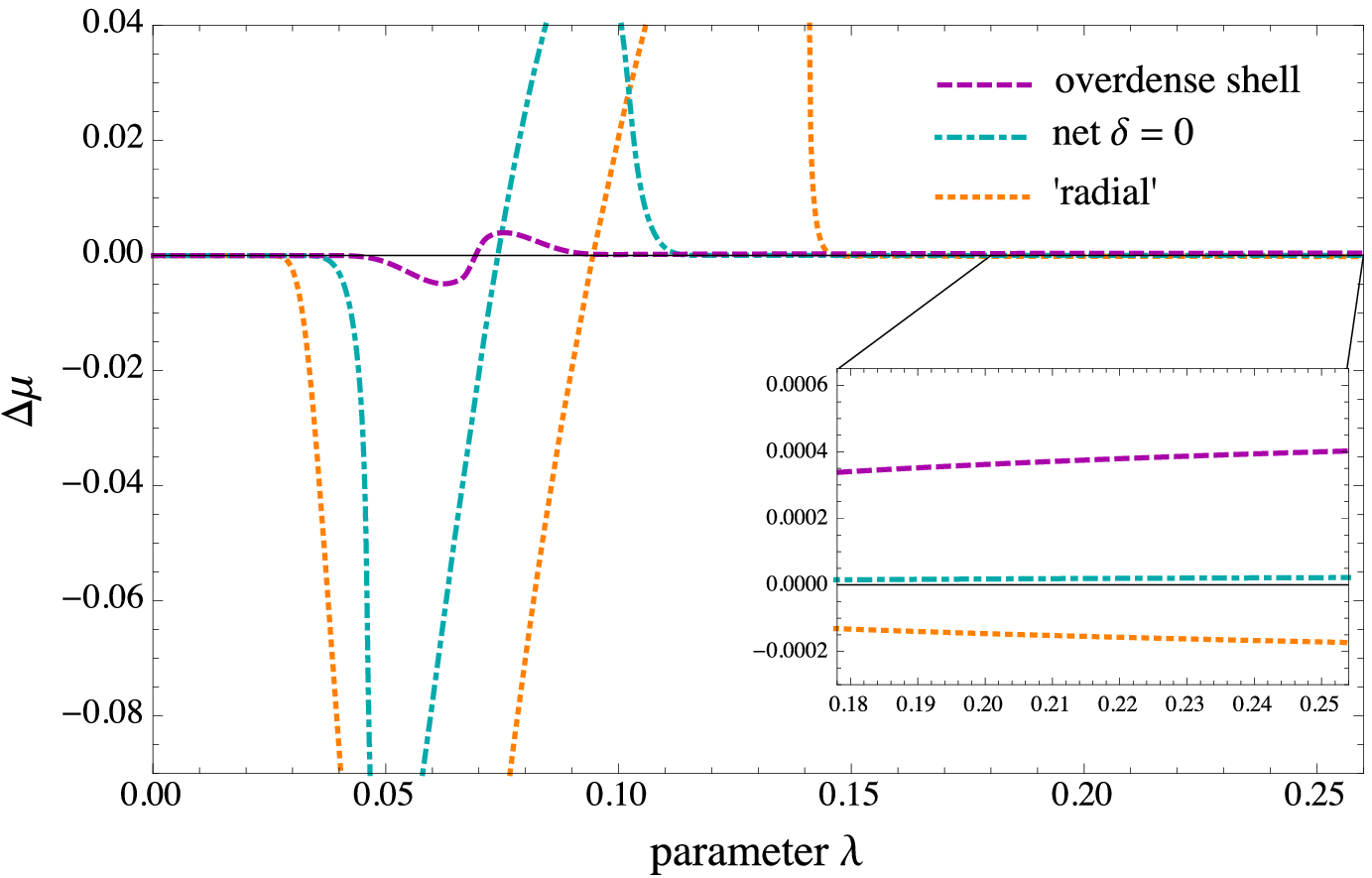}&
  \includegraphics[scale=0.6]{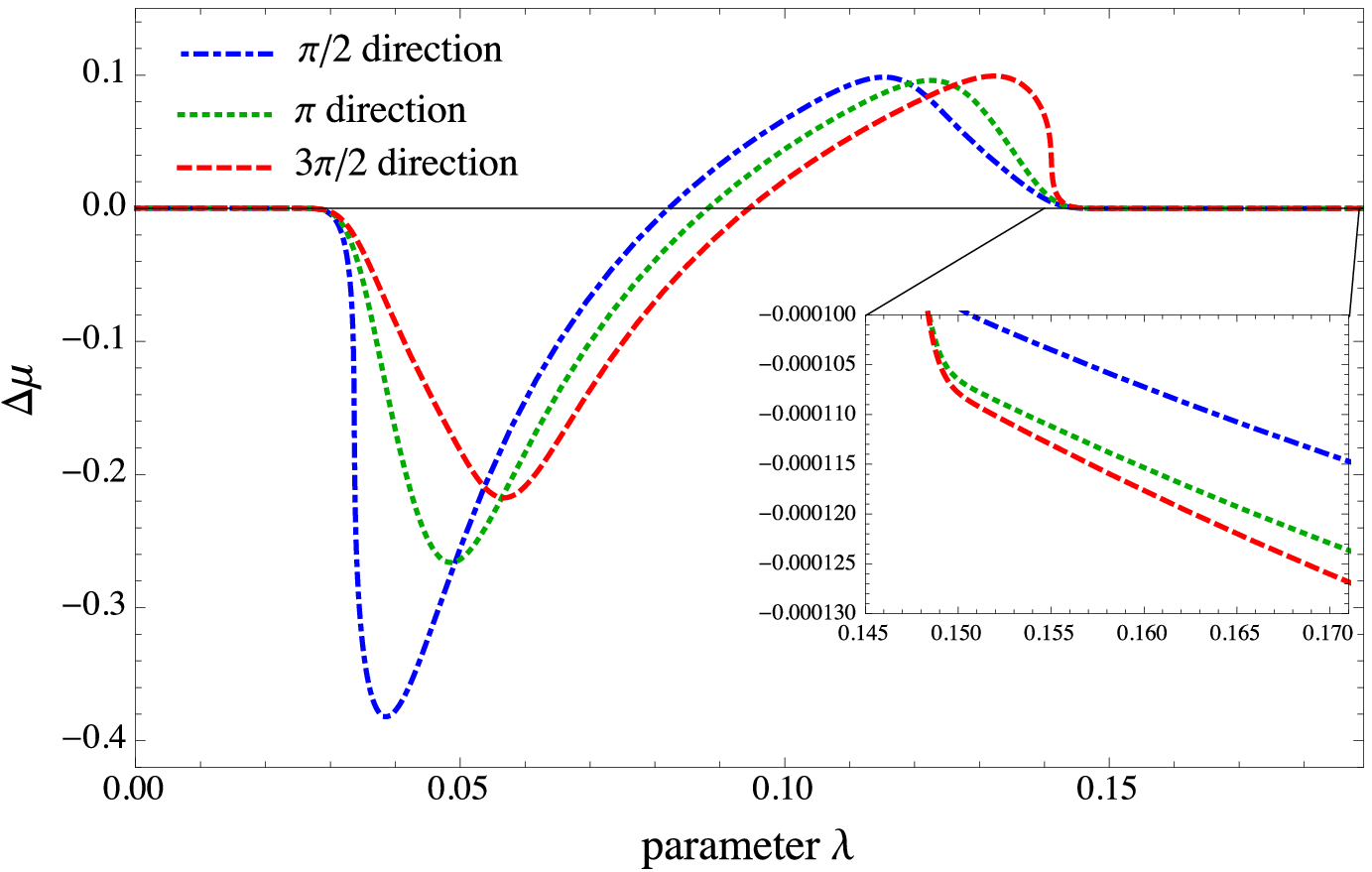}\\
\end{tabular}
\caption{LEFT: Deviations in distance modulus for light beams propagating along different paths through a single Szekeres hole as a function of the affine parameter. Depicted are a demagnified `radial' path (dotted, orange), as well as two other `non-radial' paths that exhibit the magnification possibilities for beams crossing a hole. As seen in the zoom panel, the `radial' path ends with $\Delta\mu<0$, and so is demagnified due to spending the maximum possible time in the underdense interior of the hole. The dashed (magenta) curve shows the shift in $\mu$ for a light beam that has essentially just grazed the outer overdense compensating shell, and it is therefore magnified ($\Delta\mu>0$) relative to the FLRW background. The dot-dashed (cyan) curve represents a path for which the net density encountered by the beam is effectively zero, or $\delta=0$, leading to $\Delta\mu\approx 0$. The light in this case enters the structure at an angle, passes first through the overdense shell, spends some time in the underdense region, and finally exists again through the outer shell in such a way that the overall over- and underdensities encountered cancel out.
RIGHT: $\Delta\mu$ for three particular `radial' paths through the same Szekeres hole at left, where the curves are labeled by their azimuthal position (in the spherical coordinates of the surrounding FLRW spacetime) after exiting the hole. Both the $\pi/2$ and $3\pi/2$ direction paths traverse the hole along the same line, but the $\pi/2$ direction exits the structure through the maximum density of the overdense shell, while the $3\pi/2$ direction exits through the minimum. The $\pi$ direction defines a different but still radial path through the structure that is orthogonal to the (single) path described by the other two directions. The resulting $\Delta\mu$ values are similarly negative among the three cases due to the maximal time spent within the underdense region. However, the additional spread of the curves, as shown in the inset, arises due to the evolution of the structure as the beams cross it, so that different paths see effectively different lenses.}
\label{fig:mu_1hole_Szek}
\end{figure}

In the left panel of Fig. \ref{fig:mu_1hole_Szek}, we show the shift in the distance modulus $\Delta\mu$ for particular paths through a single Szekeres hole. We plot against the affine parameter $\lambda$ instead of the redshift to see more clearly the differences between the curves. The source redshifts corresponding to the maximum parameter values are 0.11 and 0.085 for the left and right panels, respectively. We define $\Delta\mu$ to be the difference between $\mu$ for a given redshift in an FLRW universe (with $\Lambda$CDM parameters) and the value obtained in the Swiss-cheese model for the same redshift; i.e., $\Delta\mu=\mu_\mathrm{FLRW}-\mu_\mathrm{SC}$. The figure illustrates that different paths through the same structure result in different $\Delta\mu$, depending on how much time the light beam spends in the overdense and underdense regions. Magnification ($\Delta\mu>0$) occurs when the light just grazes through the overdense shell, avoiding the underdense interior, and the dashed (magenta) curve represents this case. The dot-dashed (cyan) curve is for a beam that has crossed the hole in such a way that the total overdensity and underdensity encountered effectively cancel, resulting in $\Delta\mu\approx 0$. The final possibility shown is for when the beam propagates along a `radial' path through the structure so that it spends the majority of its time in the underdense region. Demagnification ($\Delta\mu<0$) occurs in this case, as the largest amplitude dotted (orange) curve shows.

These results are consistent with expectation that $\Delta\mu$ essentially reflects the integrated density encountered by the light beam. This includes the fact that the structure evolves during the time it takes light to cross it, and so there should be small differences in $\Delta\mu$ for the same path across the hole when traversed in opposite directions. The extreme example of this for the density distribution we have chosen is when the beam enters the hole through the maximum overdensity peak and exits through the minimum---and vice versa. The maximum $\delta$ available to the beam that enters at the minimum is larger than for that of the opposite direction, since clustering leads to enhancement of the peak over time. The right panel of Fig. \ref{fig:mu_1hole_Szek} illustrates these possibilities, where we have labeled the curves according to their azimuthal position (relative to spherical coordinates of the surrounding FLRW spacetime) after exiting the structure, just before reaching the observer. All three paths are essentially `radial,' where the dot-dashed (blue) curve represents a light beam that exits through the maximum overdensity peak, while the dashed (red) is for a beam that exits through the minimum. The central dotted (green) crosses the hole in an orthogonal direction to the path described by the other two curves. As the zoom panel shows, all three paths end up with $\Delta\mu<0$, since all radial paths maximize the time spent in the underdense central region. The small difference between them arises due to the evolution of the density of the structure as the beams propagate across the hole, so different paths effectively traverse different lensing structures. In the $\pi/2$ direction case, which exits through the maximum of the shell's overdensity, it receives the largest possible re-magnification (after demagnifying in the interior), so that its $\Delta\mu$ is closer to zero than the others.

Another consideration for these cases is the influence of Weyl focusing, or shearing of the beam due to tidal forces present in the anisotropic evolution of the structure. As one would expect, the magnitude of shear is larger, for example, when crossing the hole so that the density distribution as viewed from the beam is more anisotropic. For the cases in the right panel of Fig. \ref{fig:mu_1hole_Szek}, the integrated shear for the orthogonal $\pi$ direction is an order of magnitude larger than for the $\pi/2$ and $3\pi/2$ directions, similar to what was found in \cite{TIP}. However, it only leads to changes in $\Delta\mu$ that are at least two orders of magnitude smaller than $\Delta\mu$ itself, and the Ricci focusing dominates in general for an individual hole in our model.

For a single hole at low redshift, the overall departures in $\mu$ from the background are ultimately small compared to the amplitudes seen while actually passing through the holes. Stringing many such holes together randomly along a line of sight, however can result in an appreciable $\Delta\mu$, and we explore this possibility in the following sections. A final comment can be made about the lensing efficiency of the structures due to their location between the source and the observer. As with gravitational lensing in the thin lens approximation, the efficiency is maximized when the structure is approximately halfway (in angular diameter distance) between the observer and the source. We have confirmed in our model that the largest $\Delta\mu$ possible for a single hole lens is achieved using this geometry.

\begin{figure}
\centering
  \includegraphics[scale=0.7]{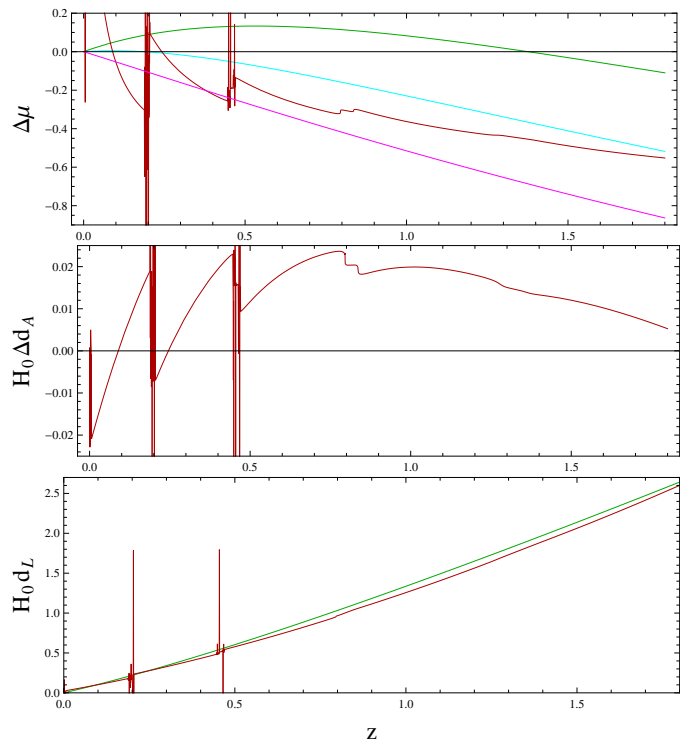}\\
\caption{Figure 13 of MKMR \cite{MKMR}, reproduced using our framework. From top to bottom, the plots show the change in distance modulus, angular diameter distance, and luminosity distance, respectively, in a Swiss-cheese universe with LT holes. The line of sight contains five aligned spherically symmetric structures, each of comoving radius approximately 120 Mpc, and the light passes radially through each. In this scenario, the inhomogeneities are extreme enough to mimic a cosmological constant density parameter of $\Omega_\Lambda=0.4$ in a background Einstein-de Sitter universe. However, other studies have shown that randomizing the impact parameters and averaging over many lines of sight effectively negates this large effect. Our goal in this work is not to obviate the need for dark energy, but rather to estimate possible bias in the distance modulus induced by non-symmetric inhomogeneities.}
\label{fig:Fig13}
\end{figure}

\subsection{Multiple holes: comparison to Swiss-cheese models with spherically symmetric holes and $\Lambda=0$}
We now consider the effect of placing multiple holes along the line of sight, and to verify our procedure, we first reproduce results from previous studies with spherically symmetric LT holes. In \cite{MKMR}, the authors (hereafter MKMR) use single-sized structures of radius approximately 120 Mpc with evacuated interiors ($\delta=-1$) and compensating mass shells. The density profiles of the structures are specified at an initial time $\bar{t}=0.2\, t_0$, where the expansion rate of the LT holes is taken to match that of the background EdS space. The overdense shell then accumulates matter from the underdense interior as the structure evolves toward the present, and the density becomes sharply peaked in the shell at $t_0$.

MKMR found that placing five such LT structures along the line of sight to distant supernovae ($z\approx1.8$) would cause an observer to erroneously conclude that he or she lives in a FLRW universe with density parameters of $(\Omega_\mathrm{m},\Omega_\Lambda)=(0.6,0.4)$. However, this universe in reality would be expanding as EdS without acceleration. In Fig. \ref{fig:Fig13}, we have reproduced the results of MKMR for this scenario, showing the effects on the distance modulus, and on the angular diameter and luminosity distances. The figure corresponds to Fig. 13 of \cite{MKMR}. Again, we have used our general Szekeres framework, reduced to its LT limit, in order to obtain the same results as MKMR, who used a different theoretical approach with a different numerical implementation.

It is of interest to note that Vanderveld et al. \cite{VFW} (hereafter VFW) showed that randomizing the impact parameters of the MKMR model effectively eliminates the large effect seen in \cite{MKMR}, and therefore such inhomogeneities cannot remove the need for dark energy. Finding that an inhomogeneous Swiss-cheese universe mimics an FLRW one with $\Omega_\Lambda=0.4$ in this case is due to considering only a special subset of possible light paths, namely perfectly radial ones. The more realistic procedure of averaging over many lines of sight with various placements of the holes does not produce a significantly different Hubble diagram from that of just the background spacetime.

More recently, Szybka in \cite{Szybka} reached the same conclusion as VFW, but by a different method. While VFW used the perturbative weak field gravitational lensing formalism and only considered magnification effects, Szybka carried out the analysis by solving the set of fully relativistic equations and evaluated the effect of shear directly. In both cases, the overall effect of properly randomized inhomogeneities on $d_A$ is negligibly small, since the average density along a typical beam path is not significantly different from the background EdS value. We have also verified that this is the case by carrying out the analysis within our framework.
\begin{figure}
\centering
\begin{tabular}{c c}
  \includegraphics[scale=0.5]{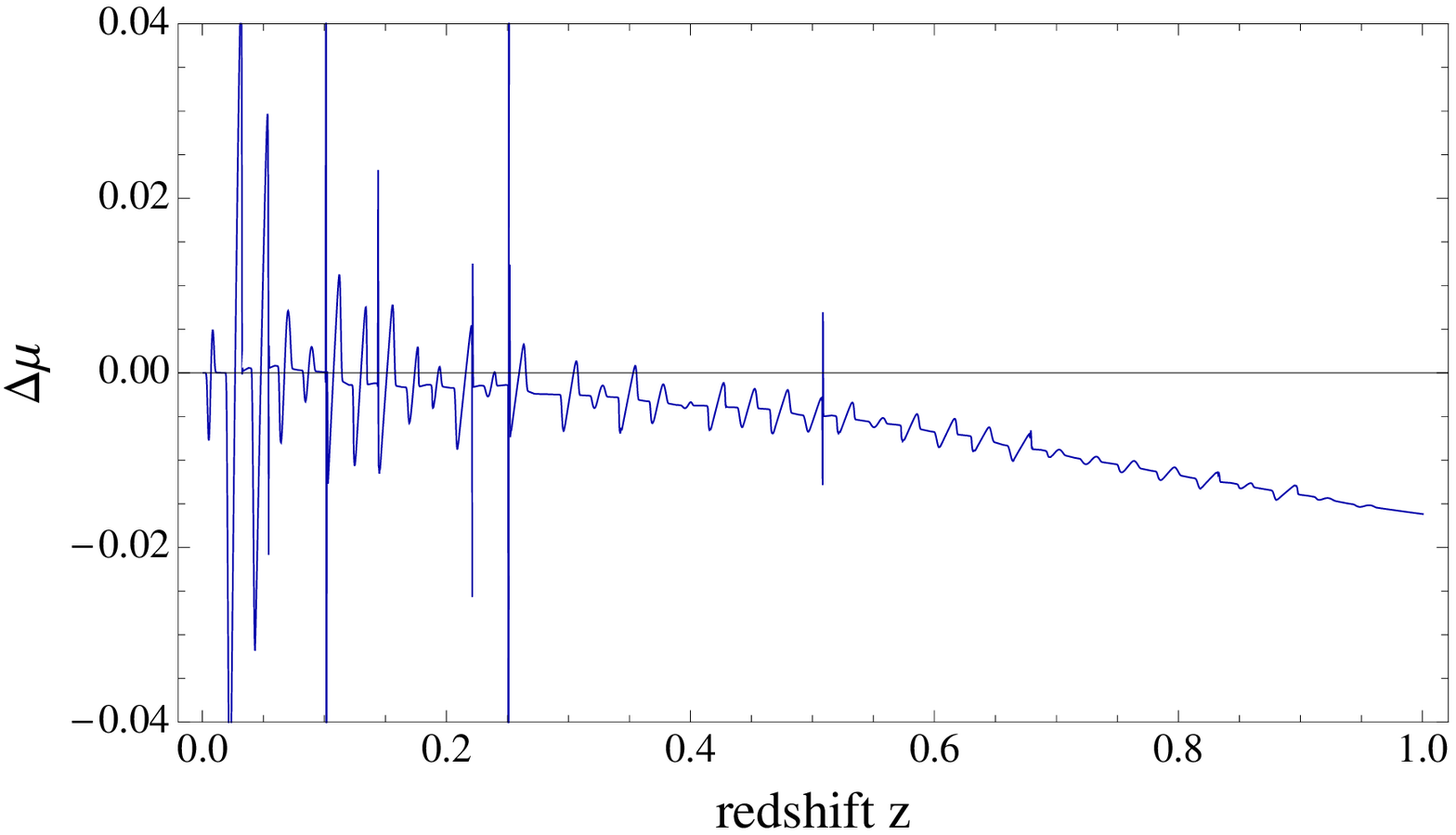}&
  \includegraphics[scale=0.48]{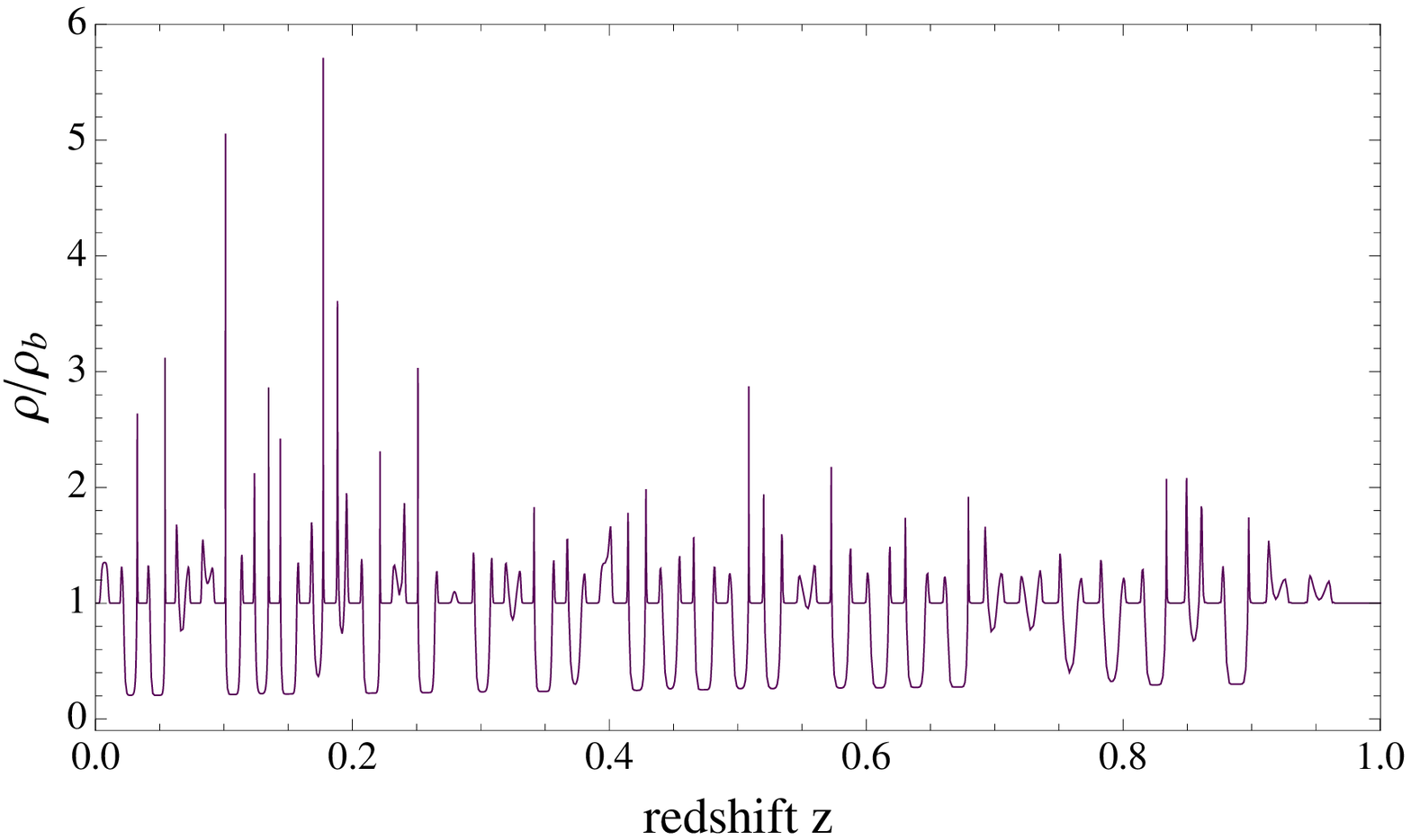}\\
\end{tabular}
\caption{LEFT: Shift in distance modulus $\Delta\mu$ relative to the $\Lambda$CDM background for a typical realization of a single beam traversing many randomized Szekeres holes. With each hole crossed, $\mu$ might increase or decrease slightly, depending on the impact parameter and orientation of the structure. RIGHT: Density (in units of the background) encountered by the same light beam shown at left as a function of redshift $z$.}
\label{fig:mu_density_38holes_Szek}
\end{figure}

\subsection{Multiple holes: a Szekeres Swiss-cheese model with $\Lambda$CDM background}\label{subsec:multiple_Szek}
For the general Szekeres Swiss-cheese case, we study the distribution of shifts in $\mu$ due to multiple anisotropic inhomogeneities along the line of sight for sources out to redshift 1.5. The left panel of Fig. \ref{fig:mu_density_38holes_Szek} is an example showing $\Delta\mu$ as a function of redshift for a light beam traversing 38 randomly oriented Szekeres holes, where small $\Lambda$CDM regions fill the spaces between them. The number of holes is dynamically determined in the code, and the source is placed at redshift 1.0 in this case. The result shown in Fig. \ref{fig:mu_density_38holes_Szek} is for one realization of this scenario and reflects the fluctuations in distance modulus for this specific realization. In the right panel of the same figure, we plot the density encountered by the beam represented at left as a function of redshift. The Szekeres structures homogenize toward the past, and, correspondingly, the maximum possible overdensities able to be seen by the beam decrease with increasing $z$. Likewise, the density contrast of the void interior increases toward zero. 

We next run a large number (1000) of random realizations for the path of light in order to study the statistics obtained from this Swiss-cheese construction and variants of it. As required by photon flux conservation \cite{Weinberg1976}, one expects a distribution with a mean of zero, while the dispersion we expect will be characteristic of the Szekeres Swiss-cheese constructions used. Such a distribution can inform us on the dispersions due to inhomogenities as modeled by these Szekeres Swiss-cheese models.

\begin{figure}
\centering
  \includegraphics[scale=0.5]{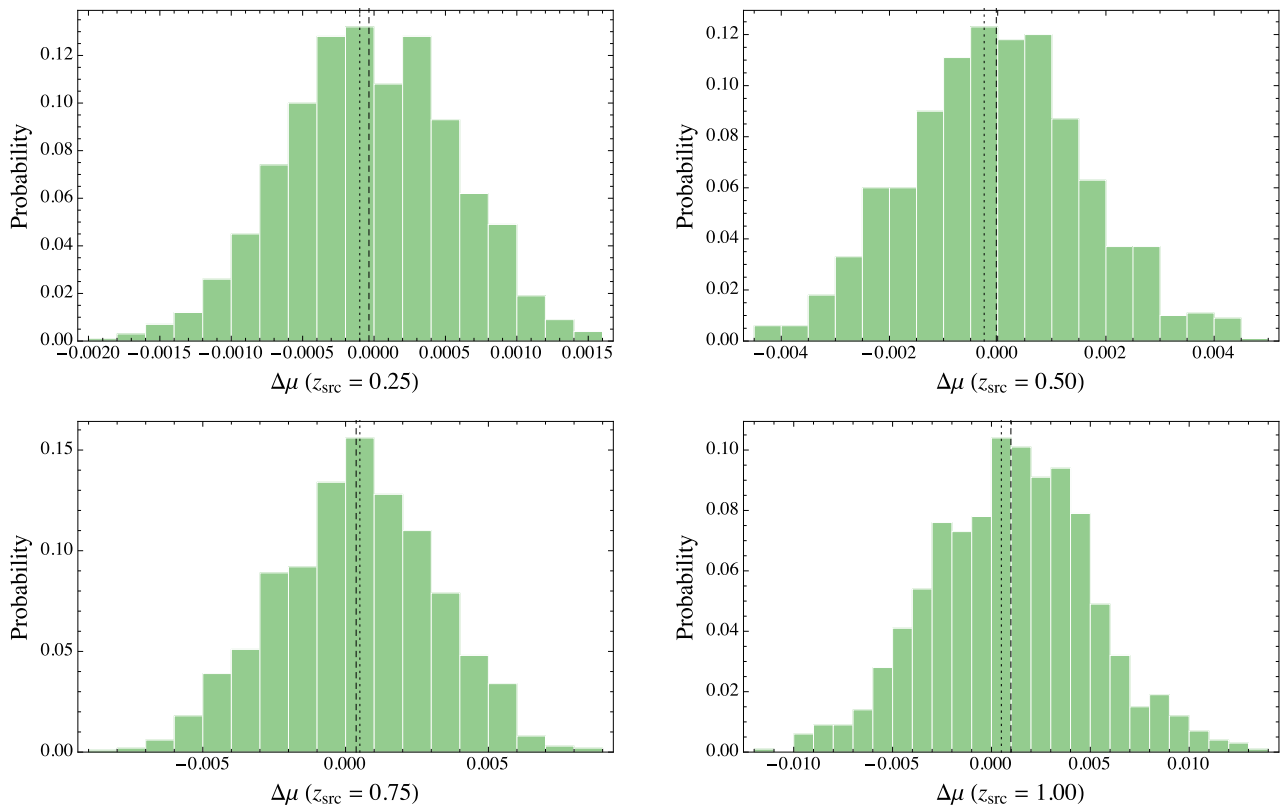}
\caption{Histograms showing the probability distributions of $\Delta\mu$ in a Szekeres Swiss-cheese model using 30 Mpc holes. Mean values are represented by a vertical dashed line and the mode by a vertical dotted line. The means are all close to zero, but the modes are not consistently toward the demagnification size, as one would expect from simulations. This seems to reflect a limitation of using such a Swiss-cheese model construction to analyze lensing properties of the real universe.}
\label{fig:30mpc_histograms}
\end{figure}

Figure \ref{fig:30mpc_histograms} shows the distribution of $\Delta\mu$ resulting from 1000 realizations of light beams emitted by sources at redshifts $\{0.25,0.5,0.75,1.0\}$. Each of the lines of sight contains randomly oriented Szekeres holes of size 30 Mpc with $\Lambda$CDM regions filling 5 Mpc spaces between them. The means $\langle\Delta\mu\rangle$ of the distribution are $\{-3.56\times10^{-5}, -2.57\times10^{-5}, 3.69\times10^{-4}, 9.83\times10^{-4}\}$, or nearly zero as expected, and the respective standard deviations are $\{5.81\times10^{-4}, 1.62\times10^{-3}, 2.74\times10^{-3}, 4.00\times10^{-3}\}$ mag. Based on the sample standard deviation, we estimate the error on the mean at, for example $z=1.0$, to be $1.27\times10^{-4}$, or about 8 standard deviations from zero. We therefore understand these constructions to exhibit a slight departure from the expected exact zero mean by flux conservation, which we see as a limitation of the model to represent the universe we intend, but this perhaps requires a further future exploration. We are also aware that it has been argued \cite{EBD,MBHE} that when caustics are included on the past light cone, one should not in fact expect the average distance in an inhomogeneous model to match that of the corresponding FLRW background, but our models do not include the formation of caustics.

\begin{figure}
\centering
  \includegraphics[scale=0.45]{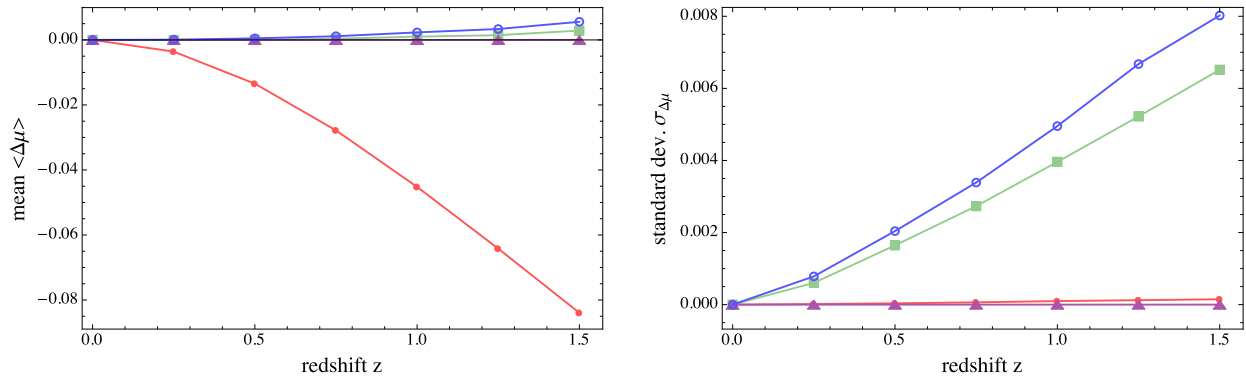}
\caption{Comparison of statistics for different Szekeres Swiss-cheese model constructions. The square data points (green) are for the 30 Mpc hole size model. The mean stays close to zero for each redshift, but the standard deviation increases up to 0.00654 mag by redshift 1.5. The triangle data points (purple) represent runs through the $\Lambda$CDM background still using the Szekeres Swiss-cheese model construction, but where the holes are all reduced to FLRW. As expected, the means are all negligible, as are the standard deviations at each redshift due to the absence of structures. The disk data points (red) are for a model where the path across each hole is taken to be random but `radial'---in other words, every crossing of a hole maximizes the time spent in the underdense region. The magnitudes of the means for this case serve as a bound on the maximum possible demagnification attainable for this model construction. The corresponding dispersion values are essentially zero as they should be.  The circles (blue) represent the case where we cycle through three different sizes of holes (30, 45, and 60 Mpc), and while its mean values remain close to zero, its standard deviations are the largest at each redshift.}
\label{fig:stats_comp}
\end{figure}

We also run simulations using hole-sizes of 45 and 60 Mpc radii, as well as cycling among the three sizes along single lines of sight. The two new larger structure sizes have density profiles similar to that of the smaller size, but we increase the maximum overdensity of the shells to 18 and 24 times the background value, respectively. Statistics for sources in $0<z\leq1.5$ are presented in Figure \ref{fig:stats_comp} for various constructions, including the 30 Mpc hole model (green squares) and the three different hole sizes model (blue circles). We see that the dispersions increase with the source redshifts and are different from one construction to another. The standard deviations at redshift one or above are in the range $0.004\le \sigma_{\Delta\mu} \le 0.008$ and remain overall below the $0.01$ mag level. We found no correlation with hole size in deviations of the means from zero, but there was a small increase in standard deviations with increasing hole size. However, the increase in $\sigma_{\Delta\mu}$ going from 45 to 60 Mpc holes was smaller than the increase going from 30 to 45 Mpc holes, and so it is not an unbounded growing effect with increasing hole size. With 60 Mpc holes (or 120 Mpc across), we feel that we have reached the size of structures like clusters and superclusters of galaxies observed in the universe, and therefore the standard deviations we find are conservative and representative of the large size inhomogeneities we aim to represent with these Szekeres Swiss-cheese constructions.

We used two limiting cases in order to confirm the dominant effect in these constructions. We first tested a case where we restricted the beam of light to travel along only aligned `radial' trajectories through the model (curve with (red) disks on Fig. \ref{fig:stats_comp}). Such a beam spends the maximum amount of time possible in the void regions. We found that it indeed exhibits a maximum of demagnification as expected, confirming that the dominant effect in our constructions is whether the ray of light spends more time in the void where it gets demagnified, or inside the shell of the structure where it gets magnified. After 1000 runs with radial beam paths, the dispersion goes to zero again as it should (curve with (red) disks). The last line with (purple) triangle data points is the FLRW limiting case with no lensing structures, and it shows zero mean and dispersion as expected. This case actually uses the full Szekeres Swiss-cheese apparatus but there the metric functions simply take on their FLRW values everywhere.

It is perhaps worth noting that one of the features resulting from using the Szekeres Swiss-Cheese model constructions here is that the distributions are not similar to those obtained from simulations \cite{Wambsganssetal,Takahashietal}. The distributions obtained there are highly skewed toward the demaginifcation side, reflecting the fact that rays get demagnified while traveling more frequently through empty space but then occasionally undergo strong magnification when they pass close to a structure. Our distributions are different and in fact show the possibility for skewness toward the magnification side as well as the demagnification side. This is perhaps due to some limitations of these constructions not being an ideal representation of real structures in the universe. This could be related to the restriction of the model to satisfy the Einstein equations without discontinuities at the boundaries (exact solutions). In other words, the matching conditions between the holes (Szekeres spacetime) and the cheese (FLRW spacetime) require compensating shells inside the holes, and this serves as a barrier to modeling the distribution of structures in the real universe.  

Also, part of the limitation of using the exact Szekeres Swiss-cheese constructions as we have done is that we do not include the effects of local structures on distance, which could be significant. Indeed, our galaxy resides within a local inhomogeneity, and as the plots in Fig. \ref{fig:mu_1hole_Szek} show, the amplitude of fluctuations in $\mu$ can be large while the light propagates through one of our structures. However, modeling the local universe using the Szekeres metric remains beyond the scope of this paper, but we have quantified here the effects on distance due to exact inhomogeneous lenses lying between the observer and a source.

Finally, we comment that no particular additional changes are expected on the distances in these Szekeres Swiss-cheese constructions for rays of light traveling all the way back to the CMB surface of last scattering. This is expected since the holes become more homogeneous in the past, meaning the effect of the inhomogeneous and anisotropic structures on light beams decreases at higher redshifts and the model reduces to essentially FLRW, as can be seen in the right panel of Fig. \ref{fig:density}. The angular diameter distance--redshift relation therefore becomes that of an FLRW model out to the CMB surface. Computations using our geodesic integration code in these models become too expensive time-wise beyond a redshift of several, but the density evolution of Fig. 2 supports this expectation.

\subsection{Multiple holes: comparison to other Swiss-cheese models}
We compare our results to the findings of Flanagan et al. in \cite{FKW} (hereafter FKW), where the authors study fluctuations in distances for two Swiss-cheese models with a $\Lambda$CDM background. The holes in their model consist of mass-compensated spherical voids, where the removed mass has been redistributed on the outer shell or within the interior of the void as galaxy halos modeled by Navarro-Frenk-White (NFW) profiles \cite{NFW}. The voids have a radius of 35 Mpc, and 90\% of the removed mass resides at the outer shell today. For sources at $z=1.5$, FKW find negligible mean distance modulus shifts of -0.0010 and -0.0013 mag with standard deviations 0.065 and 0.072 mag, respectively. The dispersions we found for our Szekeres Swiss-cheese model are smaller than theirs and the difference can be explained by the fact that the constructions differ in a number of ways. 

The FKW models contain holes of similar size to ours (and therefore contain approximately the same mass, since both are mass-compensated), but the distribution of the matter and its evolution are different. The vast majority of mass in their models is in the form of NFW profiles that do not evolve in time, whereas our Szekeres structures evolve nonlinearly but are simpler in that each hole only contains one point of maximum density in the compensating shell. Similar to the FKW models, approximately 85\% of the Szekeres hole mass resides in the outer shells today, but these shells have width on the order of 10 Mpc, while the FKW model shells are much thinner---on the order of tens of kpc. Finally, FKW use a method of computing $\Delta\mu$ due \cite{HW} that extends the standard perturbative weak lensing approach, allowing one to approximate the statistical distribution of magnifications for given void profiles. On the other hand, we use the Sachs formalism to propagate the Jacobi matrix along null geodesics and determine $d_A$, as detailed in Sec. \ref{sec:Sachs_formalism} above.

In another Swiss-cheese study, Fleury et al. \cite{FDU1,FDU2} study biases of distance measures with holes described by a Kottler spacetime. Their model specification is also quite different from ours, with the mass of the holes instead being condensed to a central opaque sphere, and where the holes lie on vertices of a regular hexagonal lattice. They find that sources are systematically demagnified---and thus appear farther---in a Swiss-cheese universe compared to a homogeneous $\Lambda$CDM universe. The authors of \cite{FDU1,FDU2} attribute the systematic bias they see to the fact that light in their models travels only through underdense regions, whereas with LT holes, the light travels through overdense regions that provide a cancellation effect. A similar cancellation therefore appears to occur in our Szekeres Swiss-cheese constructions.

\section{Summary and Conclusions}\label{sec:Conclusion}
Understanding biases in observations due to nonlinear inhomogeneous structures is crucial in an age of precise and accurate cosmology. The universe we observe is populated on all scales by voids and overdense structures that affect the characteristics of light beams as they propagate to us. In this paper, we have explored possible biases in distance determinations by studying light propagation in a Swiss-cheese universe with $\Lambda$CDM as background and holes modeled by the Szekeres metric. By construction, the entire spacetime is an exact solution of Einstein's equations, and its expansion history is identical to that of a pure $\Lambda$CDM universe. The Szekeres holes of radial size 30--60 Mpc are mass-compensated, where matter from the interior has been redistributed anisotropically around the overdense outer shells, and the underdense inner regions have $\delta=-0.8$ today to mimic voids in the real universe.

We found that the shift in distance modulus $\Delta\mu=\mu_\mathrm{\Lambda CDM}-\mu_\mathrm{SC}$ for light traversing a single Szekeres structure can be nonzero due to demagnification of the beam while traveling in the void empty space or to magnification when it goes through the overdense shell of the matter structure. There is also small additional effect due to the evolution of the structure that leads to different $\Delta\mu$ values when crossing the hole along the same line but in opposite directions. The former effect is much larger and is the dominant effect. We also verified that the effect on redshift for such inhomogeneities is negligible.

In order to model the propagation of light in the Szekeres Swiss-cheese models, we ran a large number of path realizations (1000) where the orientation of the holes and the impact parameters were randomized. We obtained probability distributions for these realizations for source redshifts of 0.25, 0.50, 0.75 and 1.00 with (nearly) zero mean, as required by photon flux conservation, and the respective dispersions in $\Delta\mu$ were $5.81\times10^{-4}$, $1.62\times10^{-3}$, $2.74\times10^{-3}$, and  $4.00\times10^{-3}$ mag. For redshifts of 1.0 and above, we found $0.004\le \sigma_{\Delta\mu} \le 0.008$ mag. Thus, the effect of the inhomogeneities as modeled by our Szekeres Swiss-cheese constructions remains below 0.01 mag, which is smaller than the intrinsic dispersion of, for example, type Ia supernova magnitudes.

It is worth commenting that the overall shape of the distributions we obtained from the Szekeres Swiss-cheese constructions were different from those derived from cosmological simulations. The distributions we found have modes (peaks) being in some cases on the demagnification side and in others on the magnification side. This is perhaps a limitation for these Swiss-cheese constructions in representing the real distribution of structures in the universe. This restriction may result from imposing compensating ovedense shells in the holes in order to satisfy the matching conditions between the Szekeres and FLRW spacetimes. This is necessary in order to avoid any discontinuity in the Einstein's equations throughout the model. 

Another limitation of using the exact Szekeres Swiss-cheese constructions as we employed them here is that the effects of local structures (inhomogeneities) on distance are not included, which could be significant. Our Fig. \ref{fig:mu_1hole_Szek} shows that the amplitude of variations in $\mu$ can be large while the beam of light propagates inside one of our structures. However, modeling the local group or cluster of galaxies using the Szekeres metric is beyond the scope of this paper.

Exploring other approaches and uses of these Szekeres Swiss-cheese constructions that would circumvent this limitation and possibly lead to different statistics remains open to future study. It will be interesting to see if that will lead to distribution more consistent with observations and other values for the dispersions.

\acknowledgments
We thank R. Biswas, P. Fleury, S. Jha, M. Kesden, L. King, M. Lavinto, M. March, and M. Wood-Vasey for useful comments during the preparation of this work. MI acknowledges that this material is based upon work supported in part by NSF under grant AST-1109667 and by the John Templeton Foundation. AP and MT acknowledge that this work was supported in part by the NASA/TSGC Graduate Fellowship program. Part of the calculations for this work have been performed on the Cosmology Computer Cluster funded by the Hoblitzelle Foundation.

\end{document}